\documentclass[aps,prx,twocolumn,showpacs,superscriptaddress,footinbib]{revtex4-1}

\usepackage[utf8]{inputenc}
\usepackage[T1]{fontenc}
\usepackage{lmodern}

\usepackage[english]{babel}
\usepackage{xcolor}
\usepackage{comment}
\usepackage{graphicx}
\usepackage{stackengine}
\usepackage{dcolumn}
\usepackage{bm}
\usepackage{bbm}
\usepackage{amssymb}
\usepackage{amsmath}
\usepackage{bbold}
\usepackage{pstricks}
\usepackage{slashed}
\usepackage{braket}
\usepackage{hyperref}
\usepackage{natbib}
\usepackage{float}
\usepackage{verbatim}
\usepackage{siunitx}
\usepackage{lipsum}
\usepackage{slashed}
\usepackage{ulem}
\usepackage[left=1.85cm,right=1.85cm,top=1.85cm,bottom=1.85cm]{geometry}

\usepackage{MnSymbol}

\definecolor{mygreen}{rgb}{0,0.5,0}
\definecolor{myblue}{rgb}{0,0,0.75}
\definecolor{mymagenta}{cmyk}{0,1,0,0.12}
\definecolor{mygray}{rgb}{0.5,0.5,0.5}

\newcommand{\Fig}[1]{Fig.~\ref{#1}}

\begin{document}
\title{Probing topological entanglement on large scales}

\author{Robert Ott}
\affiliation{Institute for Theoretical Physics, University of Innsbruck, Innsbruck, 6020, Austria}\affiliation{Institute for Quantum Optics and Quantum Information of the Austrian Academy of Sciences, Innsbruck, 6020, Austria}

\author{Torsten V. Zache}
\affiliation{Institute for Theoretical Physics, University of Innsbruck, Innsbruck, 6020, Austria}
\affiliation{Institute for Quantum Optics and Quantum Information of the Austrian Academy of Sciences, Innsbruck, 6020, Austria}

\author{Nishad Maskara}
\affiliation{Department of Physics, Harvard University, Cambridge, Massachusetts 02138, USA}

\author{Mikhail D. Lukin}
\affiliation{Department of Physics, Harvard University, Cambridge, Massachusetts 02138, USA}

\author{Peter Zoller}
\affiliation{Institute for Theoretical Physics, University of Innsbruck, Innsbruck, 6020, Austria}\affiliation{Institute for Quantum Optics and Quantum Information of the Austrian Academy of Sciences, Innsbruck, 6020, Austria}

\author{Hannes Pichler}
\affiliation{Institute for Theoretical Physics, University of Innsbruck, Innsbruck, 6020, Austria}\affiliation{Institute for Quantum Optics and Quantum Information of the Austrian Academy of Sciences, Innsbruck, 6020, Austria}

\begin{abstract}
\noindent Topologically ordered quantum matter exhibits intriguing long-range patterns of entanglement, which reveal themselves in subsystem entropies. However, measuring such entropies, which can be used to  certify topological order, on large partitions is challenging and becomes practically unfeasible for large systems. We propose a protocol based on local adiabatic deformations of the Hamiltonian which extracts the universal features of long-range topological entanglement from measurements on small subsystems of finite size, trading an exponential number of measurements against a polynomial-time evolution.
Our protocol is general and readily applicable to various quantum simulation architectures. We apply our method to various string\text{-}net models representing both abelian and non-abelian topologically ordered phases, and illustrate its application to neutral atom tweezer arrays with numerical simulations.

\end{abstract}

\maketitle

    \textit{Introduction.--} Long-range quantum correlations play an important role in various areas of physics, ranging from condensed matter~\cite{sachdev2023quantum}, to high-energy physics~\cite{polyakov1977quark} and quantum information science~\cite{kitaev2003fault}. An example is given by quantum phases with topological order (TO), which are characterized by non-local order parameters~\cite{senthil2004deconfined,mcgreevy2023generalized}. Intriguingly, the corresponding ground states feature a long-range structure of entanglement which leads to a universal correction to the area law called the topological entanglement entropy (TEE)~$S_\mathrm{topo}$~\cite{kitaev2006topological,levin2006detecting,hamma2005ground}. Recently, experimental progress for realizations of quantum states with topological entanglement has been enabled by analog quantum simulations with cold neutral atoms~\cite{semeghini2021probing,leonard2023realization,lunt2024realization}; as well as digital devices based on superconducting qubits~\cite{satzinger2021realizing,xu2024nonabelian} and trapped ions~\cite{iqbal2024non,iqbal2024topological}.

\begin{figure}[t!]
	\includegraphics[width=1.\columnwidth]{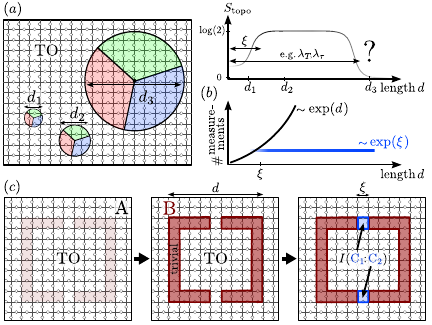}
	\caption{\textit{Protocol.} (a)~We are interested in the length scale~$d$ on which a system exhibits topological entanglement. While TEE can be visible at scales $d\gtrsim \xi$, where~$\xi$ is the (short-range) correlation length, it  might be absent on large scales $d\gg \xi$, e.g. due to thermal fluctuations (at scale~$\lambda_T$)~\cite{castelnovo2007entanglement}, or due to imperfect dynamical preparation (at scale~$\lambda_\tau$)~\cite{sahay2022quantum,giudici2022dynamical}. ($b$)~Determining TEE as suggested by Kitaev-Preskill~\cite{kitaev2006topological} and Levin-Wen~\cite{levin2006detecting} at scale $d$ involves measuring entanglement entropies on subsystems of size $\sim d^2$ (see panel (a)), which generically requires an exponential number of measurements in $d$. Instead, here we develop a protocol that requires a fixed number of measurements, set by~$\xi$. ($c$) We propose to adiabatically deform the system by trivializing region~B and leaving the region~A in a TO state, which includes the regions C$_1$, C$_2$ of size~$\xi$, separated by distance~$\sim d$. TEE at scale $d$ is extracted via the mutual information $I(\mathrm{C}_1\mathrm{:\,}\mathrm{C}_2)$.
 }
	\label{fig:Scheme}
\end{figure}

So far, however, available methods to measure the universal long-range nature of TEE and therefore certify TO, in particular over large scales, face the experimental challenge of extracting subsystem entropies of large partitions, see~\Fig{fig:Scheme}. In particular, extracting subsystem entropies without further assumptions on the underlying quantum state~\cite{elben2023randomized,daley2012measuring} generically requires a number of measurements that scales exponentially with the partition size~\footnote{Note that tomography can be made more efficient if assumptions on the quantum state can be made~\cite{joshi2023exploring}.}, and therefore such entropy measurements are currently restricted to a few qubits~\cite{satzinger2021realizing,xu2024nonabelian,iqbal2024non,iqbal2024topological,karamlou2024probing}. This prevents direct access to the spatial range of topological entanglement, which is of crucial interest for probing the stability of TO in large-scale systems \footnote{Mechanisms which can destabilize TO in these systems include thermal fluctuations~\cite{castelnovo2007entanglement,lu2020detecting} as well as finite coherence or state preparation times~\cite{sahay2022quantum,giudici2022dynamical}, and they introduce length scales beyond which TO is absent.}, see \Fig{fig:Scheme}($b$).

In this Letter, we overcome this challenge by proposing a novel protocol to efficiently probe TEE in 2D topological quantum systems. Here, we focus on \emph{robust} TO, which is energetically protected by a Hamiltonian~$H$ where both bulk and boundaries of the system are gapped. This includes the doubled topologically ordered phases with parity and time-reversal invariance represented by Levin-Wen string-net models~\cite{levin2005string}. Our protocol is based on adiabatic local deformations of $H$ which allow us to probe long-range TEE across a large spatial scale $d$ by measuring entropies on regions with finite size of order~$\sim \xi$, see~\Fig{fig:Scheme}, where $\xi$ is the (short-range) correlation length of the medium. More specifically, our protocol removes topologically trivial contributions to the entanglement entropy by decoupling parts of the subsystems indicated as regions B in \Fig{fig:Scheme}($c$), while keeping the universal TEE intact. The decoupling can be achieved in a runtime set by the minimal (finite-size) gap between the TO- and a trivial phase in region B, which scales polynomially in its size $d$ if the phases are separated by a second order transition. We thus effectively trade an exponential (in $d$) number of measurements for a poly($d$)-time evolution followed by $\exp(\xi)\sim \mathcal{O}(1)$ measurements on a reduced partition size. While our protocol is based on coherent many-body evolution and thus requires the associated coherence properties, it is largely robust to microscopic details of the system Hamiltonian. We test the protocol with numerical simulations for a class of experimentally relevant example states and illustrate its application to Rydberg atom arrays, where it can be implemented with current technology.

\textit{Protocol.--}
For concreteness, we first focus on the simplest case of $\mathbf{Z}_2$ TO, as for instance present in the ground state of the toric code Hamiltonian $H_{\mathrm{TC}}$. That is, we consider a system of spins placed on the links of an infinite two-dimensional square lattice as shown in \Fig{fig:Scheme}, and governed by the toric code Hamiltonian~\cite{kitaev2003fault}
\begin{equation}
\label{eq:TC}
    H_{\mathrm{TC}} = -\epsilon_e\sum_{v}A_v - \epsilon_m\sum_{p}B_p\, .
\end{equation}
Here $A_v = \prod_{\langle i,v\rangle} Z_i$ represent products of four Pauli-$Z$ operators acting on the links $i$ adjacent to vertices $v$, and similarly $B_p = \prod_{ i \in \Box_p} X_i$ act on the four links within plaquettes~$p$. They are weighted by magnetic and electric couplings $\epsilon_m$ and $\epsilon_e$, setting the mass scales for corresponding excitations of $m$- or $e$-anyons. The ground state $\ket{\psi_\mathrm{TC}}$ of $H_\mathrm{TC}$ is given by a coherent superposition of all states representing closed loops. For simply connected regions $R$ this results in the (von-Neumann) entanglement entropy $S_R=\mathrm{tr}\rho_R\log(\rho_R)$ of the reduced state $\rho_R$,
\begin{align}
\label{eq:SR}
    S_R = \alpha L_R - S_\mathrm{topo}\, ,
\end{align}
where $L_R$ is the boundary length of the region, $\alpha$ is a non-universal constant and $S_{\mathrm{topo}} = \log(D)$ is the TEE. $D$ represents the total quantum dimension, which is linked to the nature of~TO, and is given by $D=2$ for the case of $\mathbf{Z}_2$~TO~\cite{hamma2005ground,kitaev2006topological,levin2006detecting}. Directly extracting the value of $S_\mathrm{topo}$ from Eq.~\eqref{eq:SR} is expensive due to the area law contribution (originating from short-range fluctuations), which dominates $S_R$ at large $L_R$.

Our protocol to extract $S_\mathrm{topo}$ instead proceeds as follows: First, we introduce an external field $\propto h(t)$, which is localized only in the region~B of \Fig{fig:Scheme} and which is ramped up adiabatically. During this process the system follows the ground state of the Hamiltonian $H(t) =H_{\mathrm{TC}}  + h(t) \sum_{\ell\in \mathrm{B}} Z_\ell$. While the topological structure in the bulk (region~A) remains intact, the entanglement structure within region~B becomes trivial as loops are expelled by the external field. They are now forced to pass through the regions $\mathrm{C}_1$ and $\mathrm{C}_2$, where they lead to characteristic correlations, that can be detected efficiently. Specifically, we consider the mutual information
\begin{align}
I(\mathrm{C}_1\mathrm{:\,}\mathrm{C}_2)=S_{\mathrm{C}_1}+S_{\mathrm{C}_2}-S_{\mathrm{C}_1\mathrm{C}_2}\, ,
\end{align}
which, at the end of our protocol, is equal to the topological entropy~$I(\mathrm{C}_1\mathrm{:\,}\mathrm{C}_2)=\log(2)$, see Supplemental Material (SM). Experimentally extracting the subsystem entropies on the corresponding partitions can be achieved using standard techniques, and typically requires local control within the subsystems~\cite{elben2023randomized} or multiple copies of the system~\cite{daley2012measuring}. We note that our protocol allows us to probe presence of TO within the region enclosed by B, i.e., at scale $d$, while keeping the size of the regions~$\mathrm{C}_i$, and therefore the number of required measurements (see \Fig{fig:Scheme}($b$)), fixed at the much smaller size~$\xi^2$. 

In general, the choice of region~B depends on the scale~$\xi$ of the short-range correlations which for example result from longitudinal or transversal fields in the Hamiltonian. By choosing the regions~B to be large and smooth compared to~$\xi$, the long-distance fluctuations are described by effective quantum field theories (QFTs). The protocol drives a transition on the subregion~B, between the topological phase described by a Chern-Simons QFT~\cite{sachdev2018topological}, and a topologically trivial phase described by a Higgs QFT. If the transition is of second order (as in the case of $\mathbf{Z}_2$ TO) the gap closes polynomially in the size of region~B. Consequently, the protocol can be implemented with a runtime that is polynomial in the scale~$d$.

\begin{figure}[t!]
	\includegraphics[width=1.\columnwidth]{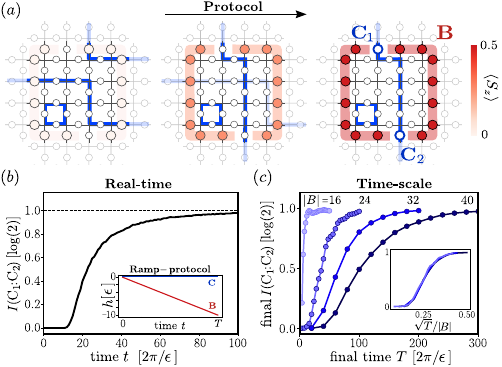}
	\caption{\textit{Protocol illustration for Toric code.} ($a$)~
  The TC ground state can be understood as a superposition of closed loops (left). By turning on a $Z$-field $h$ in region B the corresponding spins are driven into the state $\ket{\uparrow}$, and the loops are expelled from B and forced to pass through the openings $\mathrm{C}_1$ and $\mathrm{C}_2$ (center and right). This results in a non-zero mutual information $I(\mathrm{C}_1\mathrm{:\,}\mathrm{C}_2)$ reflecting the long-range entropy.
 ($b$)~The final value of $I(\mathrm{C}_1\mathrm{:\,}\mathrm{C}_2)$ yields $S_\mathrm{topo}$ in the limit of long (adiabatic) ramp times $T$. ($c$)~The ramp time depends quadratically on the boundary length~$|B|$, in accordance with the gap closing of the 1D Ising boundary theory: The curves collapse upon rescaling (inset).
 } 
	\label{fig:TC-ideal}
\end{figure}

Furthermore, there are different ways of trivializing the region~B, e.g., using longitudinal ($h(t) \sum_{\ell\in \mathrm{B}} Z_\ell$) or transversal ($g(t) \sum_{\ell\in \mathrm{B}} X_\ell$) fields. This choice of region~B corresponds to magnetic or electric boundary conditions at the interface with region~A, respectively. Then, for example, if B is a magnetic boundary our protocol measures the entropy of electric loops while being insensitive to magnetic loops (and vice versa). This is a consequence of the possibility of magnetic strings (or $m$-anyons) to condense into the region~B. It is therefore crucial that both cases are probed individually, which reveals the quantum entanglement and allows to distinguish the obtained entropies from \emph{classical} topological entropy~\cite{castelnovo2007entanglement,lu2020detecting}. In the following section, we demonstrate the protocol for $\mathbf{Z}_2$ topological matter by trivializing the region~B with longitudinal fields corresponding to the magnetic case. The electric case is dual to that and follows analogously.

\textit{Numerical Demonstration.--} In \Fig{fig:TC-ideal} we numerically demonstrate the protocol for the case of a system in the ground state of the exact toric code~\eqref{eq:TC}. As this state has correlation length $\xi=0$, it is presently sufficient to choose the region~B to be a thin, one-dimensional subregion as shown in \Fig{fig:TC-ideal}($a$). By ramping the field $h(t)$ from zero to large values (here $h=10\epsilon_e=10\epsilon_m$, as shown in panel~($b$)) at final times $T$, we drive the spins in region~B into their trivial ground state with $\langle S_z\rangle = 1/2$, as shown in the sketches in panel~($a$). For a sufficiently slow ramp evolution, we observe the building up of the topological contributions over time, resulting in a final value $I(T) = \log(2)$, while faster, non-adiabatic ramps lead to smaller values, see panel~($d$). For large systems, the required ramp times~$T$ are consistent with a polynomial scaling with the boundary length. This agrees with the linear closing of the gap in the dual Ising theory, i.e. $\Delta \propto N^{-z\nu}$, with critical exponents~$z=\nu=1$~\cite{sachdev1999quantum}.

\begin{figure}[t!]
	\includegraphics[width=.95\columnwidth]{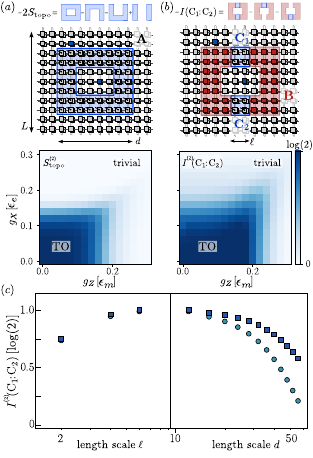}
	\caption{\textit{Detecting the length scale of topological order.} ($a$)~We consider 2D tensor-network states (A) and extract the topological (Renyi) entropy $S_\mathrm{topo}$ for various $g_X$, $g_Z$. The state phase diagram exhibits a TO and a trivial phase. ($b$)~To demonstrate that our protocol captures this phase diagram, we locally increase the value $g^Z_B$ to trivialize the region~$\mathrm{B}$ (red), and show the mutual information $I^{(2)}(\mathrm{C}_1\mathrm{:\,}\mathrm{C}_2)$. Shown system sizes [$L$,$d$,$\ell$] are [10,4,2] ($a$) and [12,4,2] ($b$).~($c$) For an incoherent anyon state the mutual information $I^{(2)}(\mathrm{C}_1\mathrm{:\,}\mathrm{C}_2)$ first increases with $\ell$, saturates to $\log(2)$ once $\ell$ surpasses the correlation length $\xi$, and subsequently decays when $d$ reaches the ``thermal'' length scale $\lambda_T$. Data is shown for $g_Z=0.15,g_X=0$. $e$-anyons are introduced by locally adapting the tensor structure (blue tensors in ($a$)), see SM.
 }
	\label{fig:TO-thermal}
\end{figure}

To go beyond the simple fixed point model with $\xi=0$ discussed above, we now illustrate our idea on a class of states with finite $\xi$. For this, we consider a class of  tensor network states which contain a topologically ordered as well as a trivial phase, and which can be efficiently computed for large system sizes~\cite{cong2024enhancing,giudici2022dynamical,cirac2021matrix,schuch2012resonating}
\begin{align}
\label{eq:PEPS-state}
    \ket{\psi} \propto e^{\sum_{\ell} (g^Z_\ell Z_\ell + g^X_\ell X_\ell)} \ket{\psi_\mathrm{TC}}\, .
\end{align}
The deformations of the TC in~\eqref{eq:PEPS-state} introduce a finite correlation length, and eventually drive a transition to a trivial phase, in analogy to the ground state of the toric code Hamiltonian perturbed by transverse and longitudinal fields. In the thermodynamic limit, these perturbations lead to a trivial topological structure for $g^Z\gtrsim 0.22$ or $g^X\gtrsim 0.22$~\cite{cong2024enhancing}, see~\Fig{fig:TO-thermal}($a$) and ($b$).
To confirm that our protocol correctly identifies these phase boundaries, we compare the TEE phase diagram of states of the form~\eqref{eq:PEPS-state} with the mutual information extracted according to our protocol, see \Fig{fig:TO-thermal}($a$),($b$). For this we mimic the state at the end of a sufficiently slow ramp with a state of the form~\eqref{eq:PEPS-state}, where we choose $g^Z_B\gg g^Z_A,g^X_A$ with subscripts A, B indicating the different regions. This allows us to efficiently compute the mutual information of the regions $\mathrm{C}_1$ and $\mathrm{C}_2$ for relatively large spatial scales $d$, $\ell$, and $L$, as shown in~\Fig{fig:TO-thermal}($b$). Here, to simplify the computation, we compute second-order Renyi entropies $S^{(2)}_{\mathrm{C}_{1/2}} = -\log[\mathrm{Tr}(\rho^2_{\mathrm{C}_{1/2}})]$, rather than von-Neumann entropies. While this changes the quantitative details of our results, it does not change their qualitative nature, as the topological Renyi entropy similarly serves as a topological invariant~\cite{halasz2012probing}. Indeed, the result of our construction agrees with the extracted topological entanglement entropy (see panel ($a$)) up to finite size corrections near the phase boundary. 

\textit{Detecting the length scale of topological entropy.--} In the following, we illustrate how our protocol detects the length scale on which topological order becomes unstable as a consequence of thermal excitations~\cite{castelnovo2007entanglement}. Specifically, we consider the following mixed state
\begin{align}
\label{eq:thermal-state}
    \rho \propto e^{g^Z \sum_{\ell} Z_\ell}e^{-\beta H_\mathrm{TC}}e^{g^Z \sum_{\ell} Z_\ell}\, ,
\end{align}
corresponding to a thermal toric-code state perturbed by local transversal fields $\sum_{\ell} g^Z_\ell Z_\ell$. We approximate this state for small temperatures, where the dominant contribution arises from zero- and two-anyon states, see~SM.

The result for the mutual information $I^{(2)}(\mathrm{C}_1\mathrm{:\,}\mathrm{C}_2)$ computed from the state $\rho$ after transforming the region~B is shown in \Fig{fig:TO-thermal}($c$) for different choices of the region B: We first increase the diameter $\ell$ of the regions $\mathrm{C}_i$, while keeping the size $d$ of region~B fixed. As expected, we obtain a value of $\log(2)$, once $d$ surpasses the short-range correlation length, $d\gtrsim \xi$. We then keep the scale~$\ell$ fixed and increase the diameter $d$ of region~B. Indeed, our protocol detects the disappearance of TEE for distances on the order of the thermal length scale, $d\sim\mathcal{O}(\lambda_T)$.

\textit{String-net models.--} As a generalization of the toric code model discussed in the previous sections, we consider different types of TO next. Specifically, we focus here on Levin-Wen string-net models~\cite{levin2005string}, such as for example based on finite abelian groups ($\mathbf{Z}_N$) or non-abelian quantum groups (e.g. $\mathrm{SO}(3)_k$), exhibiting the corresponding forms of~TO. These models involve different types of link fluxes with abelian and non-abelian fusion rules at the vertices.
\begin{figure}[t!]\includegraphics[width=1.\columnwidth]{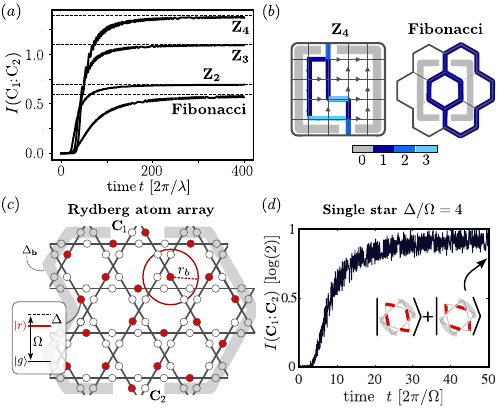}\caption{\textit{Generalized protocol.}~($a$)~Extraction of TEE with our protocol for abelian ($\mathbf{Z}_2$, $\mathbf{Z}_3$, $\mathbf{Z}_4$) and non-abelian (Fibonacci) string\text{-}net models (analogous to \Fig{fig:TC-ideal}). ($b$)~Illustration of flux quanta for individual snapshots of the given fixed-point states with the trivialized regions. ($c$)~Proposed implementation with neutral-atom tweezer arrays~\cite{semeghini2021probing}. Local control over detunings $\Delta$ of the Rydberg state $\ket{r}$ (see inset) is used to trivialize the boundary (shaded region) by forcing the atoms into their ground states $\ket{g}$. ($d$)~We numerically simulate our protocol for a small ruby lattice, see SM for detailed model parameters. At the end of the protocol, the state is dominated by the indicated snapshots (with $ |\langle\mathrm{inset}|\psi(T)\rangle|^2 \approx 0.58$). The regions $\mathrm{C}_1$ and $\mathrm{C}_2$ are chosen as the two outermost atoms respectively.} 
	\label{fig:Generalization}
\end{figure}
The Hamiltonian of a general string-net model reads
\begin{align}
     H_{\mathrm{string}\text{-}\mathrm{net}} = -\sum_{v}A_v - \sum_{p}\sum_{s} a_s B^s_p \,
\end{align}
where generalized star operators $A_v$ represent the local fusion constraints of fluxes on links connected to the vertices $v$, such that $A_v=1$ if fusion rules are fulfilled and zero otherwise. The operators $B_p^s$ add flux~$s$ on the links around a closed plaquette with weights $a_s = d_s/\sum_s d_s^2$, where $d_s$ is the quantum dimension of flux $s$, and $D =\sum_s d_s^2$~\cite{levin2005string}.

As before, we consider subregions B where we drive a phase transition to the topologically trivial phase, as for example indicated for the $\mathbf{Z}_4$ fixed point state in \Fig{fig:Generalization}($b$). In \Fig{fig:Generalization}($a$) we show the increase of the mutual information during the protocol for the cases $\mathbf{Z}_2$, $\mathbf{Z}_3$ and $\mathbf{Z}_4$, as well as for the analogous protocol for the non-abelian $\mathrm{SO}(3)_{k=3}$ string\text{-}net model hosting Fibonacci anyons. For sufficiently slow execution times, we find the topological contribution $I(\mathrm{C}_1\mathrm{:\,}\mathrm{C}_2)\rightarrow \log(D)$ in the abelian case with $D=N$, as well as $I(\mathrm{C}_1\mathrm{:\,}\mathrm{C}_2)\rightarrow -\log(D)-2\sum_{s=0}^1 (d^2_s/D)\log(d_s/D)$ with $d_0=1$ and $d_1=(1+\sqrt{5})/2$ in the non-abelian, Fibonacci~case. The topological values of the mutual information are detailed in the SM.

\textit{Application to Rydberg atom arrays.--} Our ideas are directly applicable to quantum simulators which realize TO phases in their ground states. Here we illustrate our protocol for Rydberg atom arrays, as recently used to experimentally probe the emergence of quantum spin liquid states~\cite{semeghini2021probing}. Specifically, we focus on the so-called PXP-model, $H = P[\Omega\sum_i X_i -\Delta\sum_i n_i]P$, with Rabi frequency $\Omega$ and detuning $\Delta$. $X_i= \ket{r}_i\bra{g}_i+h.c.$ and $n_i= \ket{r}_i\bra{r}_i$ act on the two levels $\ket{g}$ (ground), and $\ket{r}$ (Rydberg) of a single atom $i$, see~\Fig{fig:Generalization}($c$). The PXP-Hamiltonian operates in the fully blockaded subspace (indicated by the blockade projector $P$), which amounts to neglecting the long-range tails of the Rydberg van-der-Waals interaction beyond the blockade radius $r_b$. It serves as a toy model for analog quantum simulation experiments with Rydberg atom arrays~\cite{bernien2017probing}. On a ruby lattice, it features a $\mathbf{Z}_2$ TO phase for suitable parameter regimes~\cite{verresen2021prediction}\footnote{To stabilize TO in the presence of long-range interactions the positions of atoms are deformed with respect to the ruby lattice~\cite{verresen2021prediction,semeghini2021probing}.}.

In \Fig{fig:Generalization}($c$) we illustrate the protocol for a TO ground state with $e$-boundaries (see Ref.~\cite{verresen2021prediction} for detailed parameters). TEE is revealed as we ramp the boundary detuning $\Delta_\mathrm{b}$ of the atoms in the shaded region to force them into their ground states $\ket{g}$ (i.e. creating an $m$-region), resulting in topological correlations between the regions C$_1$ and C$_2$. This process is showcased in \Fig{fig:Generalization}($d$) for a minimal system (see inset). Over the course of the protocol the mutual information increases from initially $\approx 0$ to the value $\approx \log(2)$ at the final time~$T$, see SM for details. In general settings, we expect the ideal values to emerge for large systems as well as thickened boundaries and regions~$\mathrm{C}_i$.

\textit{Discussion.--} We have presented a scalable, polynomial-time protocol to measure the topological entanglement structure in two-dimensional quantum systems. This provides a novel tool for addressing the stability of topological quantum matter by detecting the length scales on which mechanisms such as thermal fluctuations~\cite{castelnovo2007entanglement,lu2020detecting} or state preparation defects~\cite{sahay2022quantum,giudici2022dynamical} can destabilize TO, as illustrated in~\Fig{fig:Scheme}. By effectively reducing the size of the partitions from which entropies are extracted, our protocol is both sample efficient and robust to measurement errors.

Importantly, our protocol requires the presence of a Hamiltonian, and is therefore especially suited for analog simulators realizing topological quantum matter. On the one hand, the Hamiltonian preserves the topological structure under the adiabatic deformations of the system; on the other hand, it prevents the build-up of topological long-range correlations during the protocol if absent initially. However, this also means that our procedure is sensitive to local errors in the execution of the protocol, creating pairs of anyons which could propagate to opposite sides of the trivializing region~B and therefore affect the topological structure (see SM).

Our protocol complements other techniques for learning the topological structure of quantum systems, for instance based on measurements of loop- and string order parameters~\cite{polyakov1977quark,hooft1978phase,fredenhagen1986confinement}. These non-local operators typically decay exponentially with distance, such that it becomes practically unfeasible to determine them on large scales. While this can be mitigated using decorated loop operators~\cite{cong2024enhancing}, our protocol is distinguished by the fact that it is agnositc to such microscopic details.

\begin{figure}[t!]
	\includegraphics[width=.85\columnwidth]{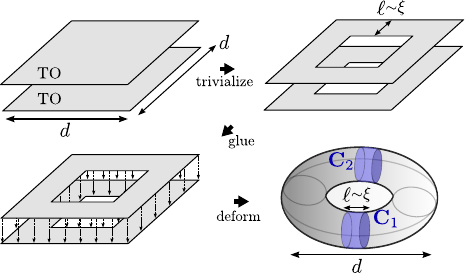}
	\caption{\textit{Glueing a torus.}~We propose to trivialize, glue and deform two sheets of topologically ordered matter as shown from left to right. TEE is efficiently extracted through the mutual information between regions $\mathrm{C}_1$ and $\mathrm{C}_2$. This construction avoids physical boundaries in the interior of the sphere (see main text). Spatial features are thickened on the scale of the correlation length~$\xi$.}
	\label{fig:wormholes}
\end{figure}

\textit{Generalizations \& Outlook.--} Our protocol can be modified in several ways. For instance, instead of adiabatically deforming the Hamiltonian,   the topologically ordered states might also directly be prepared in the presence of a trivializing field in region~B, e.g., following a generalization of the dynamical preparation protocol used in Ref.~\cite{semeghini2021probing} (and further analyzed in Refs.~\cite{giudici2022dynamical,sahay2022quantum}). Furthermore, a witness for topological, long-range correlations can in principle be obtained from measuring correlation functions of only a few local operators between $\mathrm{C}_1$ and~$\mathrm{C}_2$, e.g. in only two different bases. This is both sample efficient and avoids the need for complete tomographic control on these regions. Moreover, our protocol could be combined with other novel methods of extracting entropies~\cite{vermersch2023many,teng2024learning,zache2022entanglement,joshi2023exploring}.

Our protocol allows for further generalizations to arbitrary 2D manifolds, for instance by gluing sheets of topological matter along their boundaries as shown in~\Fig{fig:wormholes}. This can be achieved by for example slowly turning on the star and plaquette interactions between the spins at the boundaries. Such constructions avoid internal magnetic/electric boundaries, and therefore circumvent the necessity for separate measurements with different boundary conditions. Measuring the mutual information of two regions $\mathrm{C}_1$ and $\mathrm{C}_2$ of size $\mathcal{O}(\xi^2)$, the construction allows to measure TEE $2\log(2)$ on the scale~$d$ with $\mathcal{O}(\exp \xi)$ measurements, see SM. Combining time-reversed sheets can extend our protocol to chiral topologically ordered phases such as represented by fractional quantum Hall states~\cite{qi2012general,kitaev2006topological}, which have recently been realized with neutral atoms in optical potentials~\cite{leonard2023realization,lunt2024realization}.

The here obtained results are furthermore relevant beyond the 2D abelian and non-abelian topologically ordered phases discussed in this work. This includes the case of higher dimensions~\cite{grover2011entanglement} as well as other systems, such as the Yang-Mills gauge theories describing fundamental forces in particle physics~\cite{radivcevic2016entanglement}. These theories similarly feature non-local loop operators, whose long-range entanglement structure could yield further insight into, for instance, the confinement of quarks~\cite{polyakov1977quark,hooft1978phase}.

\textit{Acknowledgments.--} We thank Soonwon Choi, Simon Evered, Alexanda Geim, Byungmin Kang, and Giulia Semeghini for fruitful discussions.\\
Work in Innsbruck is funded by Horizon Europe programme HORIZON-CL4-2022-QUANTUM-02-SGA via the project 101113690 (PASQuanS2.1), the ERC Starting grant QARA (Grant No.~101041435), the EU-QUANTERA project TNiSQ (N-6001), and by the Austrian Science Fund (FWF): COE 1 and quantA. Work at Harvard University was supported by IARPA and ARO, under the Entangled Logical Qubits program (Cooperative Agreement Number W911NF-23-2-0219), DARPA ONISQ program (grant number W911NF2010021), DARPA IMPAQT program (grant number HR0011-23-3-0012), CUA (a NSF Physics Frontiers Center, PHY-1734011), NSF (grant number PHY-2012023 and grant number CCF-2313084), NSF EAGER program (grant number CHE-2037687), ARO MURI (grant number W911NF-20-1-0082), ARO (award number W911NF2320219 and grant number W911NF-19-1-0302). N.M. acknowledges support by the Department of Energy Computational Science Graduate Fellowship under award number DE-SC0021110.

\bibliographystyle{apsrev4-1} 
\bibliography{Bib_TO}

\end{document}


\title{Supplemental Material to ``Probing topological entanglement on large scales"}

\renewcommand\thefigure{SM\arabic{figure}}    
\setcounter{figure}{0}    

\maketitle

We detail the results discussed in the main text of the manuscript ``Probing topological entanglement on large scales". This Supplemental Material includes details of (A) the duality we use to perform the numerical fixed-point calculations, (B) the tensor network construction, (C) analytical calculations of entanglement entropies in the toric code, (D) the effect of local errors in the protocol, (E) the definition of the considered non-abelian string-net model and calculation of entanglement entropies, as well as (F) calculations on a minimal model of a PXP-model in a Kagome lattice.  
\subsection{Duality}
\subsubsection*{$\mathbf{Z}_2$ duality}
Here we make use of the well-known duality between a $\mathbf{Z}_2$ lattice gauge theory and a transverse Ising model in two dimensions. Specifically, the duality implements the following operator mappings
\begin{align}
    \mathrm{TC} &\rightarrow \mathrm{Ising}\nonumber \\
    B_p  &\rightarrow X_p\, , \\
     Z_{\ell_x} &\rightarrow Z_p Z_{p+y} \, ,\\
     Z_{\ell_y} &\rightarrow Z_p Z_{p+x} \, .
\end{align}
For the toric code fix-point, we do not consider additional transverse or longitudinal states, i.e. $h_\mathrm{bulk}=g_\mathrm{bulk}=0$, and the only additional terms arise from the ramp protocol with $h(t)\neq 0$, i.e.
\begin{equation}
    H = H_{\mathrm{TC}} + h(t)\sum_{i\in B}Z_i ,
\end{equation}
where $B$ again denotes the boundary region.
Hence, the Hamiltonian after the duality transform reads
\begin{align}
    H_\mathrm{Ising} = -\epsilon_m\sum_p X_p + h(t)\sum_{i=\partial B}\sum_{p\in B} Z_{p}Z_{p+i} \; . 
\end{align}
Here, the $ZZ$-terms only have support along the boundary shown in Fig.~2. Therefore, all plaquettes in the bulk commute with all other operators and can subsequently be set to the $X_p$-eigenstate with eigenvalue~$1$. The remaining evolution is reduced to the boundary problem of the one-dimensional Ising chain.
\begin{align}
    H_\mathrm{1D\,Ising} = -\epsilon_m\sum_{p\in B} X_p + h(t)\sum_{i=\partial B}\sum_{p\in B} Z_{p}Z_{p+i} \; . 
\end{align}

We simulate the remaining inhomogeneous, one-dimensional system using matrix product states with maximum bond dimension of $\chi = 64$ and a time step of $\mathrm{d}t = 0.05/\epsilon$, where $\epsilon = \epsilon_m =\epsilon_e$.

\subsubsection*{$\mathbf{Z}_N$ string-net models and Ising duality}

In the case of $\mathbf{Z}_N$, we consider a square lattice where links host degrees of freedom for $N$~local flux states $\ket{s}$, $s=0,...,N-1$, which could for instance be realized with qudits. The fusion rules enforce the sum of these fluxes to be zero at each vertex, i.e. $[\sum_{\ell}s_\ell]\mathrm{mod} N = 0$, as for example illustrated in~Fig.4($b$). For abelian groups $d_s = 1$, and hence the plaquette operators may be written as 
\begin{align}
a_s B^s_p = \frac{1}{N} (\tau_{p,\hat{x}}\tau_{p+\hat{x},\hat{y}}\tau^\dagger_{p+\hat{y},\hat{x}}\tau_{p,\hat{y}}^\dagger)^s ,
\end{align}
where the local Hilbert space is that of an $N$-state clock model with operators $\sigma$ and $\tau$ fulfilling $\sigma\tau = e^{i2\pi/N}\tau\sigma$, as well as $\tau^N=\sigma^N = 1$. Here, $p$ labels the lower left site of each plaquette $p$. $\sigma$ signals the clock state, i.e. $\sigma\ket{s} = e^{2\pi i s/N}\ket{s}$ and $\tau$ rotates the states anti-clockwise by the angle $2\pi/N$, such that $\tau \ket{s} = \ket{(s+1)\mathrm{mod}N}$.

\begin{figure*}[t!]
 \includegraphics[width=1.8\columnwidth]{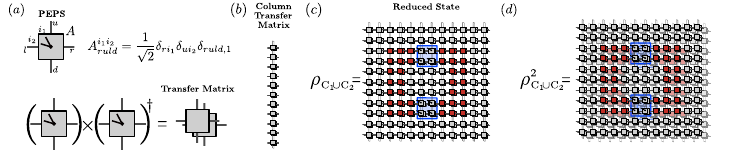} 
 \caption{\textit{PEPS construction.} ($a$) We construct the PEPS tensors $A$ (as well as $A^{(e)}$ in the presence of an $e$-anyon). To include a non-zero correlation length, we apply further matrices to the physical legs $i_1$, $i_2$ as discussed in the main text, Eq.~(4). The tensor and its conjugate are contracted along the physical legs to form a local transfer matrix. ($b$) Several local transfer matrices are contracted to form a column transfer matrix. ($c$) The column transfer matrix serves as a two-layer MPO which can be applied multiple times to form a 2D tensor network. Leaving the physical legs of the tensors in regions $\mathrm{C}_1$ and $\mathrm{C}_2$ uncontracted forms a reduced state, e.g. $\rho_{\mathrm{C}_1 \cup \mathrm{C}_2}$. The left and right boundaries are closed through two-layer boundary MPS. ($d$) To compute Renyi entropies, we multiply multiple layers of this construction and contract them along the open indices in the regions $\mathrm{C}_1$ and $\mathrm{C}_2$.}
 \label{fig:PEPS}
\end{figure*}

We propose to introduce the additional terms
\begin{align}
    \label{eq:magnetic}
    H =  H_{\mathbf{Z}_N} - h(t) \sum_{i\in B} (\sigma_i + \sigma_i^\dagger) - g(t) \sum_{i\in B} (\tau_i + \tau_i^\dagger)\, ,
\end{align}
to condense the magnetic ($h$) or electric ($g$) anyons in subregion~B.

Focusing on magnetic boundary conditions, i.e. $h\neq 0$, $g = 0$, the duality relation is given by
\begin{align}
    \mathbf{Z}_N &\rightarrow \mathrm{Clock}\nonumber \\
    B^s_p  &\rightarrow \tau_p^s\, , \\
     \sigma_{\ell_x} &\rightarrow \sigma_p^\dagger \sigma_{p+y} \, ,\\
     \sigma_{\ell_y} &\rightarrow \sigma_p^\dagger \sigma_{p+x} \, .
\end{align}
Analagous to the Ising case, the system reduces to a one-dimensional boundary problem for the fixed-point state. The Hamiltonian is
\begin{align}
    H_\mathrm{1D\,Clock} = -\frac{\epsilon_m}{N}\sum_{p\in B}\sum_{s=0}^{N-1} \tau_p^s + h(t)\sum_{i=\partial B}\sum_{p\in B} (\sigma^\dagger_{p}\sigma_{p+i} + h.c.) \; . 
\end{align}

\subsection{Tensor network calculations}

\noindent In the main text we considered the mixed state
\begin{align}
    \rho \propto e^{\sum_{\ell} g^X_\ell X_\ell}e^{-\beta H_\mathrm{TC}}e^{\sum_{\ell} g^X_\ell X_\ell}\, .
\end{align}
For instance, in the case of small temperatures $\beta\epsilon_e\gg 1$, we may expand the thermal toric code state as
\begin{align}
    e^{-\beta H_\mathrm{TC}} = \ket{\psi_\mathrm{TC}}\bra{\psi_\mathrm{TC}} &+ \frac{1}{2}e^{-2\beta\epsilon_e} \sum_{ee'} V_{ee'}\ket{\psi_\mathrm{TC}}\bra{\psi_\mathrm{TC}}V_{ee'}\nonumber\\
    &+\mathcal{O}(e^{-4\beta\epsilon_e})\, .
\end{align}
Here, $V_{ee'}$ is the unitary operator which creates two $e$-anyons at positions $e$ and $e'$. In Fig.~3 of the main text, we approximate this state with a mixture of no- and two anyon states, where one of the anyons sits in-between the regions $\mathrm{C}_1$ and $\mathrm{C}_2$. We furthermore fix the positions of the two anyons, i.e. we consider the state 
\begin{align}
    \rho = \ket{\psi_\mathrm{TC}}\bra{\psi_\mathrm{TC}} &+ \frac{1}{2}e^{-2\beta\epsilon_e}  V_{ee'}\ket{\psi_\mathrm{TC}}\bra{\psi_\mathrm{TC}}V_{ee'}\, .
\end{align}

\subsubsection*{PEPS constructions}
We use a tensor network representation of the above state to efficiently compute its entropy, where we follow Ref.~\cite{cong2024enhancing}. That is, for the toric code ground state $\ket{\psi_\mathrm{TC}}$ we consider the PEPS tensors
\begin{align}
    A^{i_1i_2}_{ruld} = \frac{1}{\sqrt{2}} \delta_{ri_1}\delta_{ui_2} \delta_{ruld,1}\, ,
\end{align}
where tensor indices $i_1$, $i_2$ describe the \emph{physical} indices, and $r$, $u$, $l$, $d$ the right, up, left, down \emph{virtual} indices, see \Fig{fig:PEPS}, which take the values $1$ and $-1$. Correspondingly, an $e$-anyon would correspond to the PEPS tensor
\begin{align}
    A^{(e),i_1i_2}_{ruld} = \frac{1}{\sqrt{2}} \delta_{ri_1}\delta_{ui_2} \delta_{ruld,-1}\, ,
\end{align}
such that flux lines start/end at the tensor. 

To construct the tenser network by contracting the tensors as shown in \Fig{fig:PEPS}: We combine the individual PEPS tensors with their conjugate to construct a local tansfer matrix. Then multiple transfer matrices are connected to produce a column transfer matrix as shown in ($b$). The left edge of the system is taken as a two-layer boundary matrix product states (MPS) and the tensor network is constructed by applying column transfer matrices as matrix product operators (MPO) onto the boundary MPS.

\subsection{Analytic calculations in the toric code}
\begin{figure}[t!]
 \includegraphics[width=.5\columnwidth]{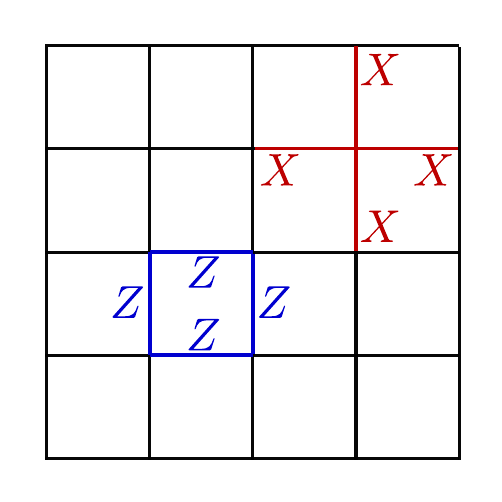} 
 \caption{\textit{2D toric code.} The toric code features spin-$1/2$ variables on the links, and stabilizers $A_s$ and $B_p$ shown in red and blue.}
 \label{fig:TC}
\end{figure}

In this section, we summarize the analytic calculations for the entanglement of the toric code ground state. In the following, we consider a two-dimensional system as discussed in the main text and shown in \Fig{fig:TC}.  
The planar system shown in \Fig{fig:TC} has a unique ground state which is fixed by the stabilizers $A_s=B_p=+1$, $\forall s,p$. For (smooth) boundary length $N$ the system has $n_\mathrm{spins}=2N(N+1)$ spin degrees of freedom. Furthermore, there are $n_P = N^2$ plaquette stabilizers and $n_S= (N+1)^2-1$ star stabilizers. Hence, after imposing the stabilizer constraints for the ground state, there are 
\begin{align}
    n_\mathrm{spins} - n_S - n_P = 0
\end{align} degrees of freedom left. The ground state on the system in  \Fig{fig:TC} is given by
\begin{align}
    \ket{\psi} = \frac{1}{2^{n_P/2}} \prod_p (1+B_p) \ket{0}\, ,
\end{align}
where $\ket{0}$ represents the vacuum with all spins pointing upwards.

\subsubsection*{Ground state entanglement}
We start by computing the ground-state entanglement with respect to the following simple bipartition
\begin{align}
\label{SM:single-connected}
\includegraphics[width=.7\columnwidth]{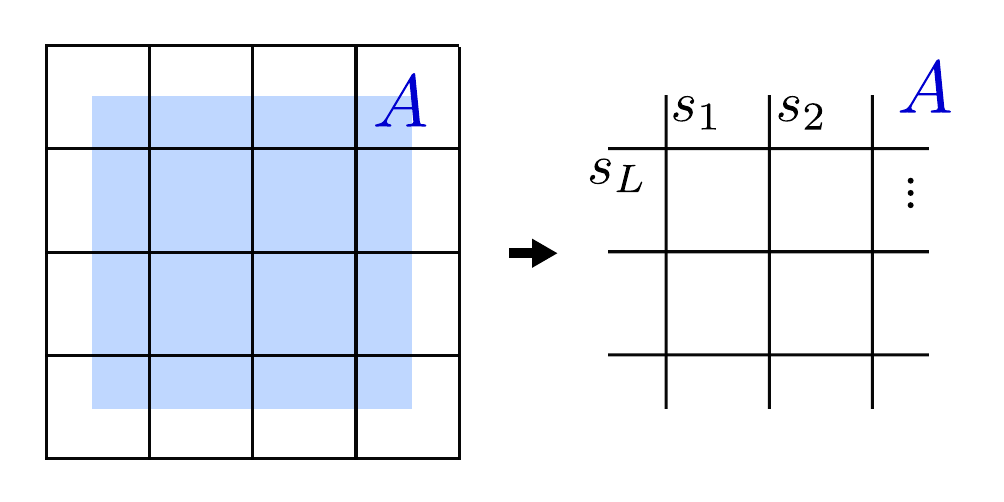} \, ,
\end{align}
where the $L=2(L_x+L_y)$ boundary spins are labelled as $s_1,...,s_L$, and the bipartition cuts across $L$ plaquettes in this case. Therefore the number of "intact" plaquettes within the subsystem $A$ is $n_{P,A} = (L_x-1)(L_y-1)$ as well as $n_{S,A} = L_xL_y$ stars. The number of spins in $A$ is given $n_{\mathrm{spins},A} =L_x(L_y+1)+L_y(L_x+1) $. That results in
\begin{align}
    n_{\mathrm{spins},A} - n_{S,A} - n_{P,A} = L-1
\end{align}
free degrees of freedom, which we label with the $L-1$ boundary spins $s_1,...,s_{L-1}$, which are fixed by the outside partition $\bar{A}$.

The reduced state is subsequently given by
\begin{align}
\rho_A&=\mathrm{tr}_{\bar{A}}\ket{\psi}\bra{\psi}\nonumber\\
&= \frac{1}{2^{L-1}}\sum_{\{s\}} \ket{\{s\}}\bra{\{s\}}\, ,
\end{align}
where $\ket{\{s\}} = \ket{s_1,...,s_{L-1}}$ labels the boundary state. Therefore, the entanglement entropy is given by
\begin{align}
    S_A &= -\mathrm{tr} \rho_A \log(\rho_A)\nonumber\\
    &= -\sum_{\{s\}} \frac{1}{2^{L-1}}\log(2^{-(L-1)})\nonumber\\
    &= (L-1)\log(2)\, ,
\end{align}
which implies $S_\mathrm{topo}=\log(2)$ in this case.

Furthermore, we consider the bipartition
\begin{align}
\includegraphics[width=.7\columnwidth]{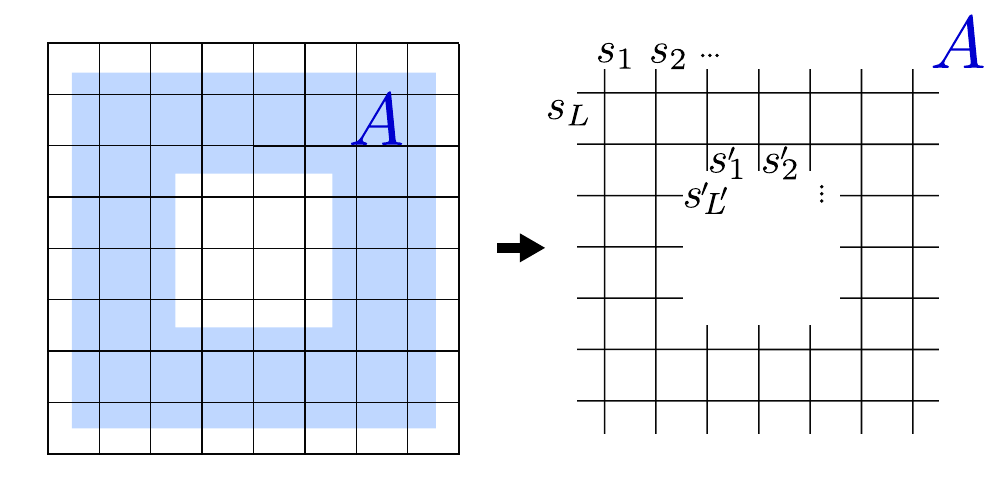} \, ,
\end{align}
which creates a "hole" in the system. Again, the number of spins is given by $n_{\mathrm{spins},A}=L_x(L_y+1)+L_y(L_x+1) - [ L'_x(L'_y-1)+L'_y(L'_x-1) ] $. The number of plaquettes is $n_{P,A} = (L_x-1)(L_y-1)- [(L'_x-1)(L'_y-1)+L']= (L_x-1)(L_y-1)- [(L'_x+1)(L'_y+1)]$; the number of stars $n_{S,A} = L_xL_y-
L'_xL'_y$. Overall that leaves us with
\begin{align}
    n_{\mathrm{spins},A} - n_{S,A} - n_{P,A} -2 = (L-1) + (L'-1)\, .
\end{align}
degrees of freedom, where we included an additional $-2$ to account for two additional constraints resulting from products of stabilizers outside of $A$, whose support lies only within $A$: the product of all stars inside the hole of $A$, and the product of all plaquettes inside the hole of $A$. The remaining configurations can be labelled as $\ket{\{s\},\{s'\}} = \ket{s_1,...,s_{L-1};s'_1,...,s'_{L-1}}$, and therefore the reduced state is 
\begin{align}
    \rho_A = \frac{1}{2^{L+L'-2}}\sum_{\{s\},\{s'\}} \ket{\{s\},\{s'\}}\bra{\{s\},\{s'\}}\, .
\end{align}
The von-Neumann entanglement entropy subsequently reads
\begin{align}
    S_A = (L+L'-2)\log(2)\, .
\end{align}
Combining the insights from both cases, yields the result of the Levin-Wen construction (shown in Fig.3(a))
\begin{align}
    S_{LW} = 2\log(2) = 2S_\mathrm{topo}\, . 
\end{align}

A third bipartition is given by
\begin{align}
\includegraphics[width=.7\columnwidth]{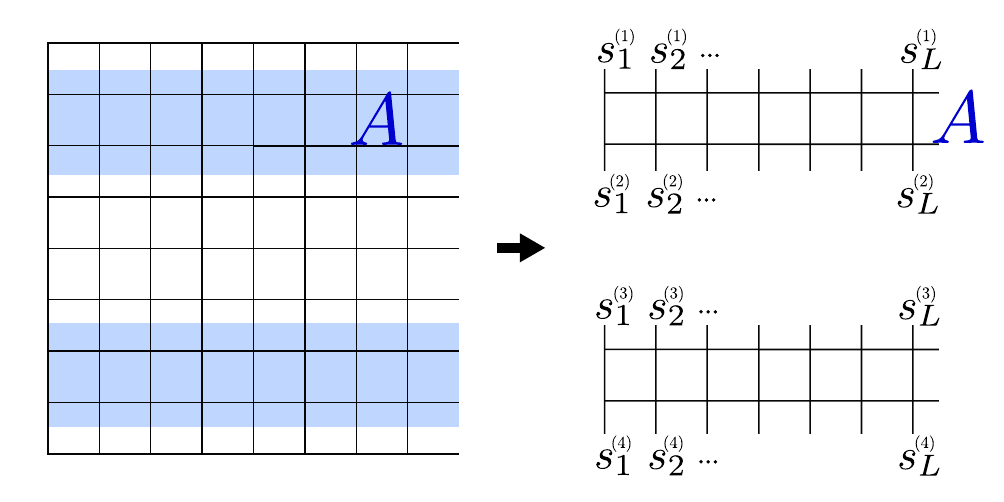} \, ,
\end{align}
with periodic boundary conditions on the right boundary, such that the system is topologically equivalent to the punctured sphere shown in Fig.5. We may again count the number of degrees of freedom. The number of spins is $n_\mathrm{spins,A} = 2 [L(W+1) +LW]$, where $W$ is the width of the strip. The number of plaquettes is $n_{P,A} = 2L(W-1)$ and the number of stars is $n_{S,A}=2LW$. Therefore, we have
\begin{align}
    n_{\mathrm{spins},A} - n_{S,A} - n_{P,A} -3 = 4L -3\, 
\end{align}
degrees of freedom, where we included an additional $-3$ for the three independent constraints that emerge from the product of: (1) all stars above $A$, (2) all stars between both disconnected parts of $A$, and (3) all plaquettes between both disconnected parts of $A$. The entanglement entropy of the partition is given by
\begin{align}
    S_A = (4L-3)\log(2)\, .
\end{align}
The result for a single disconnected region is corrspondingly given by $n_{\mathrm{spins},A_\mathrm{top}} - n_{S,A_\mathrm{top}} - n_{P,A_\mathrm{top}} -1 = 2L-1$; and therefore the mutual information between the two regions of $A$, by
\begin{align}
    I(A_\mathrm{top}:A_\mathrm{bottom}) = \log(2) \, . 
\end{align}

The relevant case for Fig.~5 is the case of a torus, which is topologically equivalent to the following bipartition
\begin{align}
\label{SM:torus}
\includegraphics[width=.7\columnwidth]{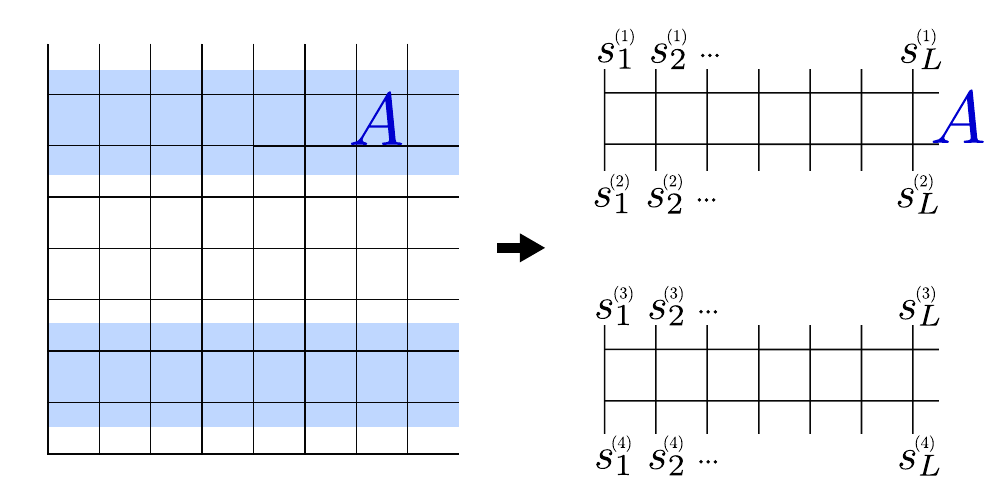} \, ,
\end{align}
with periodic boundary condition. The number of spins, stars and plaquettes is the same as above, the number of extra constraints however changes, and we get
\begin{align}
    n_{\mathrm{spins},A} - n_{S,A} - n_{P,A} -2 = 4L -2\, 
\end{align}
Combined with the result for a single disconnected region in this case ($n_{\mathrm{spins},A_\mathrm{top}} - n_{S,A_\mathrm{top}} - n_{P,A_\mathrm{top}} = 2L$), we obtain
\begin{align}
    I(A_\mathrm{top}:A_\mathrm{bottom}) = 2\log(2) \, . 
\end{align}
Here, we made explicit use of the logical loop configurations. As the structure of the large loops does not change during the gluing process, both logical Wilson and t'Hooft loops along the long torus direction are equal to one.

\subsubsection{Incoherent mixtures}
In this section, we will consider special cases of mixed states to illustrate the conceptual results for different constructions in thermal equilibrium. Specifically, the state we will consider is the following mixture
\begin{align}
\label{SM:eq:incoh-state}
    \rho = \frac{1}{2}\, \rho_\mathrm{TC} + \frac{1}{2} \,\rho_\mathrm{anyon}\, ,
\end{align}
where $\rho_\mathrm{anyon}$ will be a one- or two-anyon state.

We first consider again the first bipartition discussed in the previous section in \eqref{SM:single-connected}, i.e. a single connected region~$A$. We distinguish the case of (1) $\rho_\mathrm{anyon}$ being a single $m$ anyon inside of $A$, and (2) a pair of $e$ anyons, one inside $A$ and one outside $A$. 

In the second case, (2), the reduced state is given by
\begin{align}
    \rho_A = \frac{1}{2}\, \rho_\mathrm{TC,A} + \frac{1}{2} \,\rho_{e-\mathrm{anyon,A}}\, ,
\end{align}
where the corresponding individual reduced states are orthogonal, and given by
\begin{align}
     \rho_\mathrm{TC,A} &= \frac{1}{2^{L-1}}\sum_{\{s\}, \mathrm{even}} \ket{\{s\}}\bra{\{s\}}\,\\
     \rho_{e-\mathrm{anyon,A}}&= \frac{1}{2^{L-1}}\sum_{\{s\}, \mathrm{odd}} \ket{\{s\}}\bra{\{s\}}\, ,
\end{align}
where "even" and "odd" refer to the product $\prod_{i=1}^L s_i$, and all other plaquette and star stabilizer are implicitly taken to be $1$. The von-Neumann entropy of the partition $A$ with an $e$-anyon is
\begin{align}
    S^{(2)}_A &= \frac{1}{2}\,S_\mathrm{TC,A}+\frac{1}{2}\,S_{e-\mathrm{anyon,A}}+\log(2)\nonumber\\
    &=L\log(2)\, ,
\end{align}
where the superscript indicates case (2). Hence, the entropy receives an additional contribution $\log(2)$ from the incoherent flipping of a star stabilizer. Correspondingly, case (1) of an $m$-anyon yields
\begin{align}
     \rho_{m-\mathrm{anyon,A}}&= \frac{1}{2^{L-1}}\sum_{\{s\}, \mathrm{even}} \ket{\{s\},B_p=-1}\bra{\{s\},B_p=-1}\nonumber\\
     &\perp \rho_\mathrm{TC,A}\, ,
\end{align}
where all stabilizer except $B_p$, where $p$ is the anyon position, are implied to be $1$.
and hence, the von-Neumann entropy of the reduced state yields again
\begin{align}
    S^{(1)}_A &= \frac{1}{2}\,S_\mathrm{TC,A}+\frac{1}{2}\,S_{m-\mathrm{anyon,A}}+\log(2)\nonumber\\
    &=L\log(2)\, .
\end{align}

We next consider the bipartition of the cylinder shown in \eqref{SM:torus}. We focus on the relevant case in the thermodynamic limit: the size of the partitions $A_\mathrm{top}$ and $A_\mathrm{bottom}$ will be of order $\mathcal{\xi^2}$, the area between them on the order $d^2$ with $d\gg \xi$. Thus, we consider the case \eqref{SM:eq:incoh-state} with a single incoherent $m$-($e$-)anyon in the region in-between, and a second anyon (its partner) outside of both partitions.

We first consider case (1), a pair of incoherent $m$-anyons. The anyons commute with all stabilizer constraints of the individual partitions $A_\mathrm{top}$ and $A_\mathrm{bottom}$, i.e. their reduced states remain the same as in the ground state case before. However, their presence flips the sign of the additional plaquette constraint from the product of all plaquettes in-between $A_\mathrm{top}$ and $A_\mathrm{bottom}$. The resulting mutual information is given by
\begin{align}
    I^{(1)}(A_\mathrm{top}:A_\mathrm{bottom}) = \log(2) \, . 
\end{align}
Similarly, a pair of incoherent $e$-anyons in case (2) flips the corresponding star constraint resulting from the product of all star stabilizers in-between $A_\mathrm{top}$ and $A_\mathrm{bottom}$. The mutual information yields
\begin{align}
    I^{(2)}(A_\mathrm{top}:A_\mathrm{bottom}) = \log(2) \, . 
\end{align}
In both cases, thermal/incoherent anyons reduce the topological entropy by $\log(2)$, i.e. the mutual information of the chosen geometry distinguishes between ground state entanglement entropy and thermal classical entropy in both sectors. 

\subsection{Effect of local errors in the protocol}
In this section, we consider the effect of local errors on the trivializing region resulting from the experimental protocol. In the context of the toric code, we consider two kinds of local errors, $Z$ and $X$. Ramping up the region B allows one of the two anyons to condense, such that the protocol can be sensitive to the respective other anyon type. We expect the sensitivity of our protocol to depend on the involved time scales. Below, we distinguish between two scenarios:\\

\noindent First, if anyons propagate on time scales much shorter than the execution of the protocol, they equilibrate and add to the effective temperature of the state. Our protocol is sensitive to the temperature of the state as explained in the main text, however, if the region B is sufficiently sub-extensive the effect on the temperature can become irrelevant.\\

\noindent In the second case, when anyon propagation is sufficiently slow (compared to the thickness of the trivializing region and the ramp speed), the anyons get trapped in the trivialized region. Thus they do not influence the topological structure of the remaining bulk regions and the protocol remains unaffected.\\

In general, we expect the protocol to be robust in  both limits. In between, local errors could spread and thermalize the different TO ground states of the new geometry (formed by region B) and therefore the corresponding topological entanglement. This process in general depends on the microscopic details of the system.

As an example, we consider the case where trivialization happens for a thick region B by adding the longitudinal field $h\sum_{i\in \mathrm{B}} Z_i$ to the Hamiltonian. In this case $Z$ errors can condense into the region B and therefore have no effect. $X$ errors flip two neighboring star stabilizers within B. In the TO phase, they propagate with an effective hopping parameter set by the bulk transversal field. As $h$ increases, the anyon string obtains a string tension, thus binding the two anyons together. The resulting bound state has a mass increasing with $h$ and becomes immobile with $h\rightarrow \infty$.

\subsection{Non-abelian string-nets}
\subsubsection*{Model}
\noindent In the main text, we consider a simple non-abelian generalization of our protocol. Specifically, we consider the $\mathrm{SO}(3)_{k=3}$ string-net model hosting non-abelian Fibonacci anyons on a honeycomb lattice. Here, the links degrees of freedom allows for two flux states $\ket{s}$, $s=0,1$ and meet at three-vertices, where the fusion rules restrict the system to the following flux configurations: $(s,s',s'') = \{(0,0,0),(0,1,1),(1,0,1),(1,1,0),(1,1,1)\}$. The Hamiltonian is given by
\begin{align}
\label{eq:H-Fibonacci}
    H_\mathrm{Fib} = -\sum_v A_v - \sum_{p}\sum_{s=0}^1 a^s B^s_p  \, 
\end{align}
where $A_v = \delta_{ss's''}$ which is one if the flux configuration $(s,s',s'')$ at vertex $v$ is allowed and zero otherwise. $a^s$~are the weights of the individual plaquette operators $B^s_p$, which add flux $s$ to plaquette $p$. Here, we have $B^0_p=1$, and $a^s=d_s/D$, such that $a^0 = 1/D$, $a^1=\phi/D$, where $\phi=(1+\sqrt{5})/2$ denotes the golden ratio and $D=1+\phi^2$ is the total quantum dimension.

\subsubsection*{Topological entanglement entropy (analytics)}
As presented in the main text, the mutual information of the openings $\mathrm{C}_1$ and $\mathrm{C}_2$ does not directly yield the known form of the topological entanglement entropy $S_\mathrm{topo}=\log(D)$. However, we will demonstrate in this section that it is a topological invariant which contains information about the quantum dimensions. These in turn contain information about the nature of the topological phase. The computations in this section are mostly based on Refs.~\cite{levin2005string,levin2006detecting}, see also \cite{hahn2020generalized,bonesteel2012quantum} for more explicit expressions.

The (normalized) fixed-point state may be written as
\begin{align}
    \ket{\psi} = \frac{1}{\sqrt{D}^N} \prod_{p=1}^N (1+\phi B_p)\ket{0}\, ,
\end{align}
where $\ket{0}$ is the trivial vacuum state without flux, and $B_p$ adds flux $s=1$ around the $N$ plaquettes $p$ of the system. We are interested in computing von-Neumann entanglement entropies of the subregions shown in \Fig{fig:entanglement} in blue. To this end, we deform the system using a series of $F$-moves, where $F$'s are the unitary basis changes acting on stringnet states as
\begin{align}
 \includegraphics[width=.7\columnwidth]{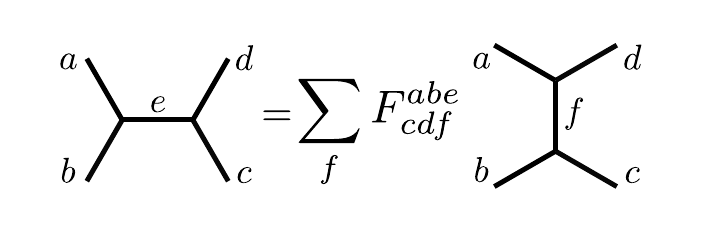} .
\end{align}
Moreover, these $F$-moves preserve the fixed-point structure, i.e. the resulting state is the fixed-point state of the new lattice configuration
\begin{align}
 \includegraphics[width=.7\columnwidth]{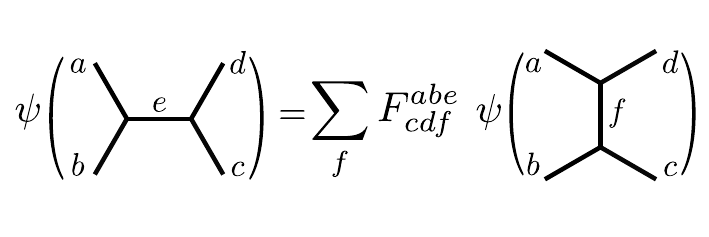} ,
\end{align}
where $\psi\left(\cdot\right)$ is the \emph{amplitude} of the fixed-point wavefunction for a specific stringnet configuration.
For the Fibonacci model, the non-trivial contributions are given by
\begin{align}
    F^{11\cdot}_{11\cdot}= \begin{pmatrix} 1/\phi && 1/\sqrt{\phi} \\1/\sqrt{\phi} && -1/\phi 
\end{pmatrix}\, ,
\end{align}
and all other $F$'s describing the local mappings between allowed stringnet configurations are equal to one. More generally, fixed-point states fulfil the following conditions
\begin{align}
 \includegraphics[width=.75\columnwidth]{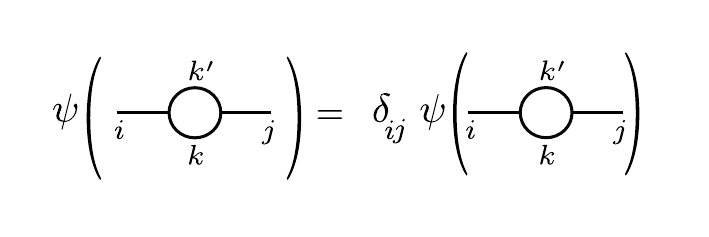} ,\\
 \includegraphics[width=.75\columnwidth]{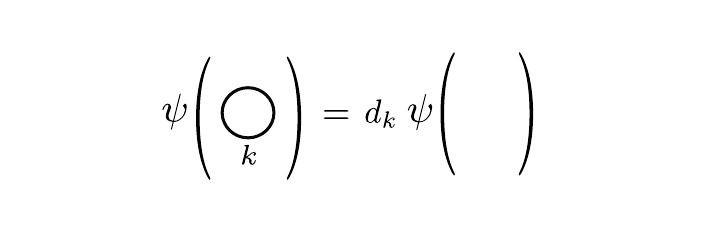} ,\\
 \includegraphics[width=.75\columnwidth]{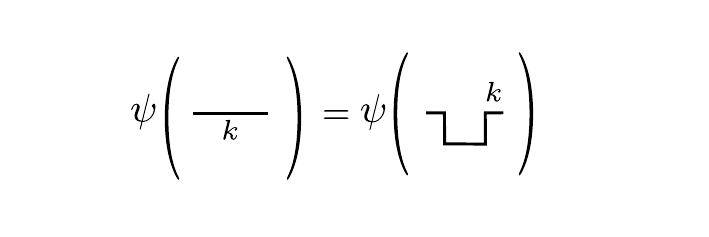} .
\end{align}

Acting with $F$-moves implies a collection of additional rules for the \emph{amplitudes} which are practical in explicit calculations. Suppressing $\psi(\cdot)$ for brevity, we have
\begin{align}
 \includegraphics[width=.75\columnwidth]{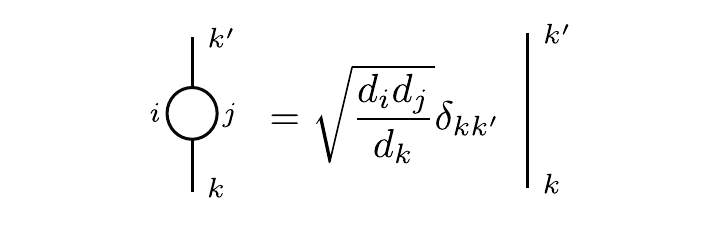} ,\\
 \includegraphics[width=.73\columnwidth]{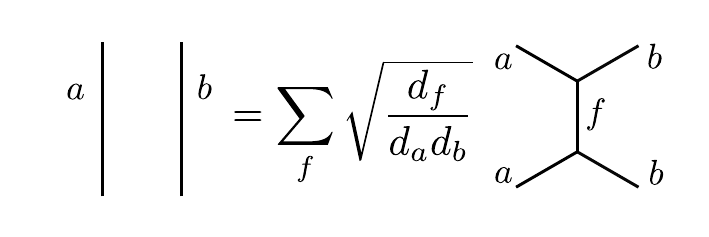} .
\end{align}

Using these rules, we may compute the reduced state and therefore the entanglement entropy of a minimal system of two plaquettes of a honeycomb lattice. The state is given by
\begin{align}
\ket{\psi}=\frac{1}{\sqrt{D}^2}\prod_p (1+\phi B^1_p)\raisebox{-6.5pt}{\includegraphics[width=.1\columnwidth]{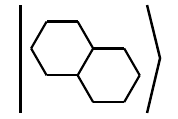}} \, .
\end{align}
Applying the above rules, we get
\begin{align}
    \ket{\psi}&=\frac{1}{D}\big(\raisebox{-6.5pt}{\includegraphics[width=.1\columnwidth]{2P0.pdf}}+\phi\raisebox{-6.5pt}{\includegraphics[width=.1\columnwidth]{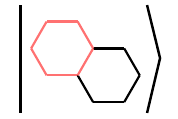}} +\phi\raisebox{-6.5pt}{\includegraphics[width=.1\columnwidth]{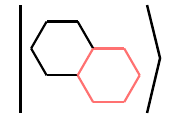}}  +\phi\raisebox{-6.5pt}{\includegraphics[width=.1\columnwidth]{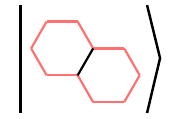}} +\phi^{\frac{3}{2}}\raisebox{-6.5pt}{\includegraphics[width=.1\columnwidth]{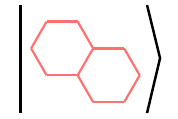}}\big)
    \, .
\end{align}

To compute the entanglement entropy of the smallest subsystem, we split the link in the center connecting the two plaquettes into a part $A$ and two more parts which belong to the remaining system $B$. Tracing out $B$, we get the following reduced state
\begin{align}
    \rho_A = \frac{1}{D} ( \ket{0}\bra{0} + \phi^2 \ket{1}\bra{1})\, ,
\end{align}
where $\ket{s}$ refers to flux state $s$ of the remaining link. The corresponding entanglement entropy is given by
\begin{align}
    S_A &= -\mathrm{tr} \rho_A \log(\rho_A)\nonumber\\
    &= -j\log(D) - n\sum_k \frac{d^2_k}{D}\log(\frac{d_k}{D})\, ,
\end{align}
which agrees with the Levin-Wen formula~\cite{levin2006detecting} for boundary length $n=2$ and for $j=1$ connected boundary components.

\begin{figure}[t!]
	\includegraphics[width=1.\columnwidth]{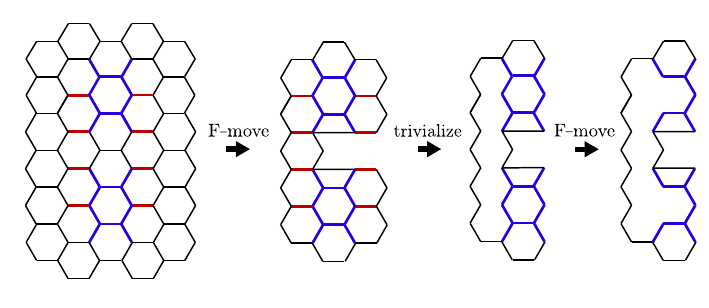}
	\caption{\textit{Computation of entanglement in extended fixed-point states.} Different steps of the computation involve $F$-moves and trivialization.}
	\label{fig:entanglement2}
\end{figure}

Following this example, we move to another example which illustrates the conceptional aspects of the general calculation. We consider a fixed-point state on the system shown in \Fig{fig:entanglement2}. Fixed-point states on any lattice configuration can be deformed as shown in the first step using the $F$-moves. The protocol trivializes the dark red links by forcing them into the zero flux state, and subsequently they can be removed. The remaining fixed-point state can be written as
\begin{align}
\ket{\psi}&=\frac{1}{D^{\frac{5}{2}}} \sum_{i,k_1,...,k_4} d_i \prod_{j=1}^4 \left( d_{k_j}\right)\raisebox{-31pt}{\includegraphics[width=.2\columnwidth]{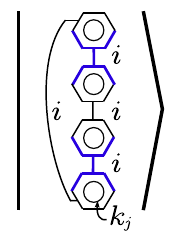}}\, .
\end{align}
Using this and the previous fusion rules, the state can be simplified to
\begin{align}
    \ket{\psi}&=\frac{1}{D^{\frac{5}{2}}} \sum_{i,k_1,...,k_4} d_i \prod_{j=1}^4 \left( d_{k_j}\right)\sum_{k'_1,...,k'_4} \prod_{j}\sqrt{\frac{d_{k'_j}}{d_{k_j}d_i}}\raisebox{-31pt}{\includegraphics[width=.2\columnwidth]{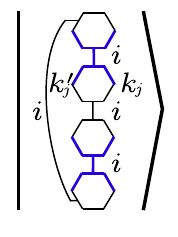}}\nonumber\\
    &=\frac{1}{D^{\frac{5}{2}}} \sum_{i,\{k_j\},\{k'_j\}} d_i  \prod_{j}\sqrt{\frac{d_{k'_j}d_{k_j}}{d_i}}\ket{i,\{k_j\},\{k'_j\}}\nonumber\\
    &=\frac{1}{D^{\frac{5}{2}}} \sum_{i,\{k_j\},\{k'_j\}} d_i^{-1}  \prod_{j}\sqrt{d_{k'_j}d_{k_j}}\ket{i,\{k_j\},\{k'_j\}} \, ,
\end{align}
where we abbreviated the state as $\ket{i,\{k_j\},\{k'_j\}}$. As before, the state is normalized
\begin{align}
    \mathrm{tr}\rho &=  \frac{1}{D^{5}} \sum_{i,\{k_j\},\{k'_j\}} d_i^{-2}  \prod_{j}d_{k'_j}d_{k_j} \delta_{k_jk'_ji}\nonumber\\
     &=  \frac{1}{D^{5}} \sum_{i,\{k_j\}} d_i^{2}  \prod_{j}d^2_{k_j} \nonumber\\
     &=1\, .
\end{align}
Here we used the fusion constraint
\begin{align}
    d_id_j = \sum_k d_k \delta_{ijk}\, .
\end{align}
Let us compute the reduced state of the blue partition $C_1C_2$ after link splitting
\begin{align}
    \rho_{12}&=\frac{1}{D^5} \sum_{i,\{k_j\},\{k'_j\}}   \frac{\prod_{j}d_{k'_j}d_{k_j}}{d_i^{2}} \ket{i,\{k_j\},\{k'_j\}}\bra{i,\{k_j\},\{k'_j\}}\, .
\end{align}
From this state, we get the von-Neumann entropy as
\begin{align}
    S_{12}&= \log(D^5)- \sum_{i,\{k_j\},\{k'_j\}}   \frac{\prod_{j}d_{k'_j}d_{k_j}\delta_{k_jk'_ji}}{d_i^{2}D^5} \log\left(\frac{\prod_{l}d_{k'_l}d_{k_l}}{d_i^{2}} \right)\nonumber\\
    &=\log(D^5)- \sum_l\sum_{i,\{k_j\},\{k'_j\}}   \frac{\prod_{j}d_{k'_j}d_{k_j}\delta_{k_jk'_ji}}{d_i^{2}D^5} \log\left(d_{k'_l}d_{k_l} \right)\nonumber\\ 
    &\qquad- \sum_{i,\{k_j\},\{k'_j\}}   \frac{\prod_{j}d_{k'_j}d_{k_j}\delta_{k_jk'_ji}}{d_i^{2}D^5} \log\left(\frac{1}{d_i^{2}} \right)\nonumber\\ 
    &=\log(D^5)- \sum_l\sum_{i,k_l,k'_l}   \frac{d_i^3d_{k'_l}d_{k_l}\delta_{k_lk'_li}}{d_i^{2}D^2} \log\left(d_{k'_l}d_{k_l} \right)\nonumber\\ 
    &\qquad- \sum_{i,\{k_j\}}   \frac{d_i^4\prod_{j}d^2_{k_j}}{d_i^{2}D^5} \log\left(\frac{1}{d_i^{2}} \right)\nonumber\\ 
    &=\log(D^5)- \sum_l\sum_{i,k_l,k'_l}   \frac{d_i^3d_{k'_l}d_{k_l}\delta_{k_lk'_li}}{d_i^{2}D^2} \log\left(d_{k'_l}d_{k_l} \right)\nonumber\\ 
    &\qquad- \sum_{i}   \frac{d_i^2}{D} \log\left(\frac{1}{d_i^{2}} \right)\, .
\end{align}
We calculate the remaining part of the expression separately. We get
\begin{align}
   - \sum_l\sum_{i,k_l,k'_l}&   \frac{d_i^3d_{k'_l}d_{k_l}\delta_{k_lk'_li}}{d_i^{2}D^2} \log\left(d_{k'_l}d_{k_l} \right)\nonumber\\
   &=-\sum_l\sum_{i,k_l,k'_l}   \frac{d_id_{k'_l}d_{k_l}\delta_{k_lk'_li}}{D^2} \log\left(d_{k_l} \right)\nonumber\\
    &\qquad-\sum_l\sum_{i,k_l,k'_l}   \frac{d_id_{k'_l}d_{k_l}\delta_{k_lk'_li}}{D^2} \log\left(d_{k'_l} \right)\nonumber\\
    &=- \sum_l\sum_{k_l}   \frac{d^2_{k_l}}{D} \log\left(d_{k_l} \right)\nonumber\\
    &\qquad-\sum_l\sum_{k'_l}   \frac{d^2_{k'_l}}{D} \log\left(d_{k'_l}\right)\, .
\end{align}
The final expression for the entanglement entropie therefore is
\begin{align}
    S_{12}&= 5\log(D) - 8 \sum_{k}   \frac{d^2_{k}}{D} \log\left(d_{k}\right) + 2 \sum_{i}   \frac{d_i^2}{D} \log\left(d_i^{2} \right)\nonumber\\
    &= -\log(D) - 8 \sum_{k}   \frac{d^2_{k}}{D} \log\left(\frac{d_{k}}{D}\right) + 2 \sum_{i}   \frac{d_i^2}{D} \log\left(\frac{d_i}{D} \right)\, .
\end{align}

Again, to compute the mutual information, we look at the subsystem $C_1$. The relevant state on such a system is analogously given by
\begin{align}
    \ket{\psi}&=\frac{1}{D^{\frac{3}{2}}} \sum_{i,k_1,k_2} d_i \prod_{j=1}^2 \left( d_{k_j}\right)\sum_{k'_1,k'_2} \prod_{l=1}^2\sqrt{\frac{d_{k'_l}}{d_{k_l}d_i}}\raisebox{-25pt}{\includegraphics[width=.17\columnwidth]{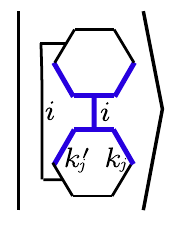}}\nonumber\\
    &=\frac{1}{D^{\frac{3}{2}}} \sum_{i,\{k_j\},\{k'_j\}} d_i  \prod_{j}\sqrt{\frac{d_{k'_j}d_{k_j}}{d_i}}\ket{i,\{k_j\},\{k'_j\}}\nonumber\\
    &=\frac{1}{D^{\frac{3}{2}}} \sum_{i,\{k_j\},\{k'_j\}}   \prod_{j}\sqrt{d_{k'_j}d_{k_j}}\ket{i,\{k_j\},\{k'_j\}} \, .
\end{align}
For the von-Neumann entropy we obtain
\begin{align}
    S_{1}&= \log(D^3)- \sum_{i,\{k_j\},\{k'_j\}}   \frac{\prod_{j}d_{k'_j}d_{k_j}\delta_{k_jk'_ji}}{D^3} \log\left(\prod_{l}d_{k'_l}d_{k_l} \right)\nonumber\\
    &= 3\log(D)- 4 \sum_{k}   \frac{d^2_{k}}{D} \log\left(d_{k}\right)\nonumber\\
    &= -\log(D)- 4 \sum_{k}   \frac{d^2_{k}}{D} \log\left(\frac{d_{k}}{D}\right)\, ,
\end{align}
which is equal to the corresponding result $S_2$. Combining everything, the mutual information is given by
\begin{align}
    I(\mathrm{C}_1:\mathrm{C}_2) &= -S_{12}+S_1+S_2\nonumber\\
    &= -\log(D)  - 2 \sum_{i}   \frac{d_i^2}{D} \log\left(\frac{d_i}{D} \right)\, ,
\end{align}
as stated in the main text. We anticipate that this formula extends to larger system sizes and, moreover, to the case away from the fixed-point state if all structures are thickened as discussed in the main text. That is, because the construction of the mutual information subtracts all short-range entanglement of $\mathrm{C}_1$ and $\mathrm{C}_2$ with their respective local environment.

\subsubsection*{Numerics (exact diagonalization)}
To demonstrate the measurement of a topological mutual information, we employ exact diagonalization of the Hamiltonian \eqref{eq:H-Fibonacci} on the system of five plaquettes shown in Fig.~4(b) of the main text (dashed links are guides for the eye and have been removed for simplicity). Starting from the fixed-point state, i.e. the ground state of $H_\mathrm{Fib}$, we dynamically trivialize the link degrees of freedom along the red boundary region (B) with the Hamiltonian  
\begin{align}
    H = H_\mathrm{Fib} + \Delta \sum_{i\in \mathrm{B}}\ket{s=1}_i\bra{s=1}_i\, ,
\end{align}
which introduces an energy penalty for flux, i.e.~$s=1$. We compute the entanglement entropy by splitting the links in the entanglement regions $\mathrm{C}_1$ and $\mathrm{C}_2$ (the openings of the red region B) and embedding the states on both subsystems into the product Hilbert space $\mathcal{H}_{\mathrm{C}_1\cup\mathrm{C}_2}\otimes \mathcal{H}_{\overline{\mathrm{C}_1\cup\mathrm{C}_2\cup B}}$, and similarly for the individual entanglement entropies $S_{\mathrm{C}_1}$. Subsequently we compute entanglement entropies in standard fashion.

\subsection{Rydberg atom array (PXP)}
In Fig.4($d$) of the main text we showcase the application of our protocol for a single star of a two-dimensional ruby lattice of Rydberg atoms. The calculation was performed with exact diagonalization in the restricted Rydberg blockade subspace. That is, the Hilbert space is the product space $\{\ket{g},\ket{r}\}^{\otimes N}$ restricted to the subspace where no two atoms on links which meet at a shared vertex are simultaneously excited to the Rydberg state~$\ket{r}$. This restriction is indicated by the blockade projector~$P$.

The Hamiltonian is chosen as given in the main text. Here, we additionally differentiate between the detuning $\Delta$ in the bulk, and the \emph{boundary} detuning $\Delta_\mathrm{b}$ on the atoms which sit along the outer boundary of the star. We first diagonalize the Hamiltonian with parameters $\Omega/2\pi =1$, $\Delta/2\pi =4$, $\Delta_\mathrm{b}/2\pi =2$. We then evolve the system in real time and linearly increase the boundary detuning of the atoms indicated by the grey region in Fig.4($d$) by $\Delta_\mathrm{b}(T)/2\pi = \Delta_\mathrm{b}(0)/2\pi +10$ during the ramp evolution. We extract the mutual information from the remaining boundary atoms at times $t$, which yields the result shown in Fig.4($d$). 

\begin{figure}[t!]
\flushright
 \includegraphics[width=1.\columnwidth]{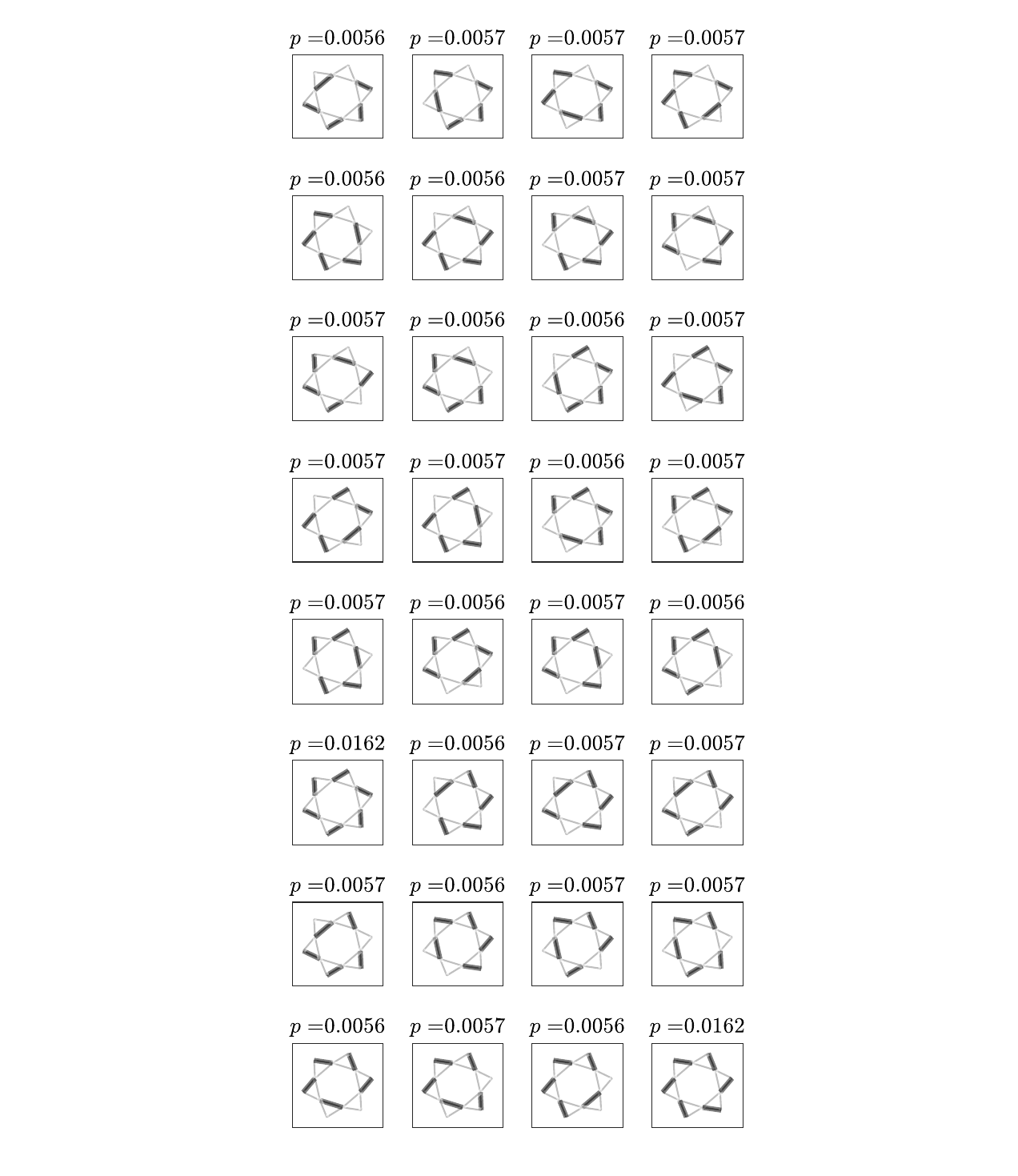} 
 \caption{\textit{Single star initial state.} We consider the initial state to the single-star Hamiltonian with parameters $\Omega/2\pi =1$, $\Delta/2\pi =4$, $\Delta_\mathrm{b}/2\pi =2$. The shown configurations have the highest weights $p$. Rydberg excitations are shown in dark grey.}
 \label{fig:initial-state-single-star}
 \centering
 \includegraphics[width=.49\columnwidth]{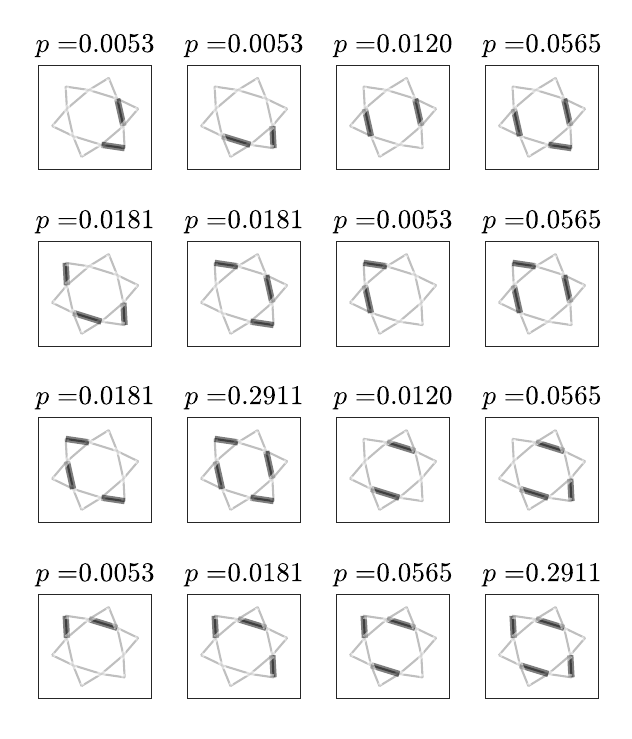} 
 \caption{\textit{Single star final state of protocol.} We consider the final state after the protocol for the single-star Hamiltonian with initial parameters $\Omega/2\pi =1$, $\Delta/2\pi =4$, $\Delta_\mathrm{b}/2\pi =2$. The shown configurations have the highest weights $p$. Rydberg excitations are shown in dark grey.}
 \label{fig:final-state-single-star}
\end{figure}

For the small system size considered here, the definition of quantum phases is ambiguous and hence we do not distinguish between a topological spinliquid state and a valence bond solid (VBS). The protocol would, however, distinguish the two in an extended system as shown in Fig.4($c$) with thickened boundary region~B analogous to Figs.1 and 3. Therefore the chosen parameters are fine tuned to obtain an initial state with resembles resonating dimer coverings, see~\Fig{fig:initial-state-single-star}. The final state after the protocol is shown~\Fig{fig:final-state-single-star}.

\clearpage

\bibliographystyle{apsrev4-1} 
\bibliography{Bib_TO}